\documentstyle[aps]{revtex}

\def \z {{(0)}}
\def \f {{(1)}}

\begin{document}
\twocolumn[\hsize\textwidth\columnwidth\hsize\csname
@twocolumnfalse\endcsname

\title{On the Stability of the Einstein Static Universe}

\author{John D. Barrow$^1$,
George F.R. Ellis$^2$, Roy Maartens$^{2,3}$, Christos G.
Tsagas$^2$ }

\address{~}

\address{$^1$DAMTP, Centre for Mathematical Sciences, Cambridge
University, Cambridge~CB3~0WA, UK}

\address{$^2$ Department of Mathematics \& Applied Mathematics,
University of Cape Town, Cape~Town~7701, South~Africa}

\address{$^3$ Institute of Cosmology \& Gravitation, University of
Portsmouth, Portsmouth~PO1~2EG, UK}

\address{~}

\maketitle

\begin{abstract}

We show using covariant techniques that the Einstein static
universe containing a perfect fluid is always neutrally stable
against small inhomogeneous vector and tensor perturbations and
neutrally stable against adiabatic scalar density inhomogeneities
so long as $c_{\rm s}^2>{1\over5}$, and unstable otherwise. We
also show that the stability is not significantly changed by the
presence of a self-interacting scalar field source, but we find
that spatially homogeneous Bianchi type~IX modes destabilise an
Einstein static universe. The implications of these results for
the initial state of the universe and its pre-inflationary
evolution are also discussed.

\end{abstract}

\vskip2pc]

\section{Introduction}

The possibility that the universe might have started out in an
asymptotically Einstein static state has recently been considered
in~\cite {EM}, thus reviving the Eddington-Lema\^{\i}tre
cosmology, but this time within the inflationary universe context.
It is therefore useful to investigate stability of the family of
Einstein static universes. In 1930, Eddington~\cite{E} showed
instability against spatially homogeneous and isotropic
perturbations, and since then the Einstein static model has been
widely considered to be unstable to gravitational collapse or
expansion. Nevertheless, the later work of Harrison and Gibbons on
the entropy and the stability of this universe reveals that the
issue is not as clear-cut as Newtonian intuition suggests.

In 1967, Harrison~\cite{har} showed that all physical
inhomogeneous modes are oscillatory in a radiation-filled Einstein
static model. Later, Gibbons~\cite{G} showed stability of a
fluid-filled Einstein static model against {\em conformal} metric
perturbations, provided that the sound speed satisfies $c_{{\rm s}
}^{2}\equiv dp/d\rho >{\frac{1}{5}}$.

The compactness of the Einstein static universe, with an
associated maximum wavelength, is at the root of this
``non-Newtonian" stability: the Jean's length is a significant
fraction of the maximum scale, and the maximum scale itself is
greater than the largest physical wavelength. Here we generalize
Gibbons' results to include general scalar, vector, and tensor
perturbations and a self-interacting scalar field. We also show
that the Einstein static model is unstable to spatially
homogeneous gravitational wave perturbations within the
Bianchi~type IX class of spatially homogeneous universes.

Consider a Friedmann universe containing a scalar field $\phi $
with energy density and pressure given by
\[
\rho _{\phi }=\frac{1}{2}\dot{\phi}^{2}+V(\phi ),~p_{\phi
}=\frac{1}{2}\dot{ \phi}^{2}-V(\phi ),
\]
and a perfect fluid with energy density $\rho $ and pressure
$p=w\rho $, where $-{\frac{1}{3}}<w\leq 1$. The cosmological
constant $ \Lambda $ is absorbed into the potential $V$. The fluid
has a barotropic equation of state $p=p(\rho )$, with sound speed
given by $c_{{\rm s} }^{2}=dp/d\rho $. The total equation of state
is $w_{{\rm t}}=p_{{\rm t} }/\rho _{{\rm t}}=(p+p_{\phi })/(\rho
+\rho _{\phi })$.

Assuming no interactions between fluid and field, they separately
obey the energy conservation and Klein-Gordon equations,
\begin{eqnarray}
\dot{\rho}+3(1+w)H\rho &=&0,  \label{cons} \\
\ddot{\phi}+3H\dot{\phi}+V^{\prime }(\phi ) &=&0.  \label{kg}
\end{eqnarray}
The Raychaudhuri field equation
\begin{equation}
{\frac{\ddot{a}}{a}}=-{\frac{8\pi G}{3}}\left[
{\frac{1}{2}}(1+3w)\rho +\dot{ \phi}^{2}-V(\phi )\right] ,
\label{Ray}
\end{equation}
has the Friedmann equation as a first integral,
\begin{equation}
H^{2}={\frac{8\pi G}{3}}\left[ \rho
+{\frac{1}{2}}\dot{\phi}^{2}+V(\phi ) \right] -{\frac{K}{a^{2}}},
\label{Friedmann}
\end{equation}
where $K=0,\pm 1$, and together they imply
\begin{equation}
\dot{H}=-4\pi G\left[ \dot{\phi}^{2}+(1+w)\rho \right]
+{\frac{K}{a^{2}}}. \label{b1}
\end{equation}

\section{The Einstein static universe}

The Einstein static universe is characterized by $K=1$,
$\dot{a}=\ddot{a}=0$ . It is usually viewed as a fluid model with
a cosmological constant that is given a priori as a fixed
universal constant, and this is the view taken in previous
stability investigations~\cite{E,har,G}. However, in the context
of inflationary cosmology, where the Einstein static model may be
seen as an initial state, we require a scalar field $\phi $ as
well as a fluid, and $ \Lambda $ is not an a priori constant but
determined by the vacuum energy of the scalar field. The vacuum
energy is determined in turn by the potential $ V(\phi )$ of the
scalar field, which is given by some physical model. An initial
Einstein static state of the universe arises if the field starts
out in an equilibrium position:
\begin{equation}
V^{\prime }(\phi _{0})=0\,,~~\Lambda =8\pi G\,V(\phi_0)\,.
\end{equation}

We will first discuss the simple case where $V$ is trivial, i.e.,
a flat potential. A general Einstein static model has
$\dot{\rho}=0=V'=\ddot{\phi}$ by Eqs.~(\ref{cons})--(\ref{b1}),
and satisfies
\begin{eqnarray}
{\frac{1}{2}}(1+3w)\rho _{0}+\dot{\phi}_{0}^{2} &=&V_{0},
\label{es1} \\ (1+w)\rho _{0}+\dot{\phi}_{0}^{2} &=&{\frac{1}{4\pi
Ga_{0}^{2}}}. \label{es2}
\end{eqnarray}
If the field kinetic energy vanishes, that is if
$\dot{\phi}_{0}=0$, then by Eq.~(\ref{es2}), $(1+w)\rho _{0}>0$,
so that there must be fluid in order to keep the universe static;
but if the static universe has only a scalar field, that is if
$\rho _{0}=0$, then the field must have nonzero (but constant)
kinetic energy~\cite{BB,EM}. Qualitatively, this means that there
must be effective total kinetic energy, i.e. $\rho _{{\rm
t}}+p_{{\rm t} }>0$, in order to balance the curvature energy
$a^{-2}$ in the absence of expansion or contraction, as shown by
Eq.~(\ref{b1}).

Equations~(\ref{es1}) and (\ref{es2}) imply
\begin{eqnarray}
w_{{\rm t}}&=& -{\frac{1}{3}}\,, \\ w_\phi &=&-{\frac{1}{3}}\left[
{\frac{V_0+(1+3w)\rho_0/2}{V_0- (1+3w)\rho_0/6 }}\right]\,.
\label{es3}
\end{eqnarray}
It follows that $w_\phi\leq -{\frac{1}{3}}$ since
$w>-{\frac{1}{3}}$ and $ \rho_0\geq0$. In other words, the
presence of matter with (non-inflationary) pressure drives the
equation of state of the scalar field below $-{\frac{1}{3 }}$. If
the field has no kinetic energy, then by Eqs.~(\ref{es1}) and
(\ref {es2}), its equation of state is that of a cosmological
constant,
\begin{equation}
\dot{\phi}_0=0~\Rightarrow~ w_\phi=-1\,.
\end{equation}
If there is no fluid ($\rho_0=0$), then the field has maximal
kinetic energy,
\begin{equation}
\rho_0=0~\Rightarrow~ w_\phi=-{\frac{1}{3}}=w_{{\rm t}}\,.
\end{equation}
In this case, the field rolls at constant speed along a flat
potential, $ V(\phi)=V_0$. Dynamically, this pure scalar field
case is equivalent to the case of $w=1$ pure fluid with
cosmological constant. This can be seen as follows: equating the
radii $a_0$ of the pure-field and fluid-plus-$\Lambda$ cases,
using Eq.~(\ref{es2}), implies $V_0=(1+w)\rho_0$, by
Eq.~(\ref{es1}). This then leads to $w_\phi=-(3+5w)/(5+3w)$, by
Eq.~(\ref{es3}). However, we know that $w_\phi=-1$ for the
fluid-plus-$\Lambda$ case, and equating the two forms of $w_\phi$
leads to $w=1$.

Thus there is a simple one-component type of Einstein static
model, which may be realised by perfect fluid universes with
cosmological constant (the pure scalar field case is equivalent to
the $w=1$ fluid model). This is the case considered by Eddington,
Harrison and Gibbons. The general case, of most relevance to
cosmology, involves a scalar field with nontrivial potential
$V(\phi )$, so that the Einstein static model is an initial state,
corresponding to an equilibrium position ($V_{0}'=0$). This
general case is a two-component model, since in addition to the
scalar field, a fluid is necessary to provide kinetic energy and
keep the initial model static.

\section{Inhomogeneous perturbations of the fluid model}

We consider first the effect of inhomogeneous density
perturbations on the simple one-component fluid models. Density
perturbations of a general Friedmann universe are described in the
1+3-covariant gauge-invariant approach by $\Delta =a^{2}{\rm
D}^{2}\rho /\rho $, where ${\rm D}^{2}$ is the covariant spatial
Laplacian. The evolution of $\Delta $ is given by~\cite {BDE}
\begin{eqnarray}
&&\ddot{\Delta}+\left( 2-6w+3c_{{\rm s}}^{2}\right)
H\dot{\Delta}+\left[ 12\left( w-c_{{\rm s}}^{2}\right)
{\frac{K}{a^{2}}}\right.  \nonumber \\ &&~\left. {}+4\pi G\left(
3w^{2}+6c_{{\rm s}}^{2}-8w-1\right) \rho +(3c_{ {\rm
s}}^{2}-5w)\Lambda \right] \Delta  \nonumber \\ &&~{}-c_{{\rm
s}}^{2}{\rm D}^{2}\Delta -w\left( {\rm D}^{2}+3{\frac{K}{a^{2}}
}\right) {\cal E}=0\,,  \label{ddotDi}
\end{eqnarray}
where ${\cal E}$ is the entropy perturbation, defined for a
one-component source by
\begin{equation}\label{e}
p{\cal E}=a^2{\rm D}^2p-\rho c_{\rm s}^2\Delta\,.
\end{equation}
For a simple (one-component perfect-fluid) Einstein static
background model, ${\cal E}=0$ and Eq.~(\ref {ddotDi}) reduces to
\begin{equation}
\ddot{\Delta}_{k}=4\pi G(1+w)\left[ 1+(3-k^{2})c_{{\rm
s}}^{2}\right] \rho _{0}\Delta _{k}\,,  \label{EddotnDi}
\end{equation}
where we have decomposed into Fourier modes with comoving index
$k$ (so that ${\rm D}^{2}\rightarrow -k^{2}/a_{0}^{2}$). It
follows that $\Delta _{k}$ is oscillating and non-growing, i.e.
the Einstein static universe is stable against gravitational
collapse, if and only if
\begin{equation}
(k^{2}-3)c_{{\rm s}}^{2}>1\,.  \label{Egsc1}
\end{equation}
In spatially closed universes the spectrum of modes is discrete, $
k^{2}=n(n+2)$, where the comoving wavenumber $n$ takes the values
$ n=1,2,3\dots $~in order that the harmonics be single-valued on
the sphere~\cite{lif,har}. Formally, $n=0$ gives a spatially
homogeneous mode ($k=0$), corresponding to a change in the
background, and the violation of the stability condition in
Eq.~(\ref{Egsc1}) is consistent with the known result~\cite{E}
that the Einstein static models collapse or expand under such
perturbations. This follows from the Raychaudhuri
equation~(\ref{Ray}).

For the first inhomogeneous mode ($n=1$), the stability condition
is also violated, so that the universe is unstable on the
corresponding scale if there is initial data satisfying the
constraint equations. However this is a gauge mode which is ruled
out by the constraint equations ($G_{0i}=8\pi GT_{0i}$). This mode
reflects a freedom to change the 4-velocity of fundamental
observers. (For multiple fluids with different velocities, one can
in principle have isocurvature modes with $n=1$ that are not
forced to zero by the Einstein constraint equations.) The physical
modes thus have $n\geq 2$. Stability of all these modes is
guaranteed if
\begin{equation}
c_{s}^{2}>{\frac{1}{5}}\,.  \label{st}
\end{equation}
This condition was also found by Gibbons~\cite{G} in the
restricted case of conformal metric perturbations; we have
generalized the result to arbitrary adiabatic density
perturbations. Harrison~\cite{har} demonstrated stability for a
radiation-filled model and instability of the dust-filled model,
which are included in the above result.

Thus {\it an Einstein static universe with a fluid that satisfies
Eq.~(\ref {st}) (and with no scalar field), is neutrally stable
against adiabatic density perturbations of the fluid for all
allowed inhomogeneous modes}. The case $w=1$ covers the pure
scalar field (no accompanying fluid) models. When $c_{{\rm
s}}^{2}>0$ the stability of the model is guaranteed by
Eq.~(\ref{Egsc1}) for all but a finite number of modes. The
universe becomes increasingly unstable as the fluid pressure
drops. Clearly, a dust Einstein static universe ($c_{{\rm
s}}^{2}=0$) is always unstable. In that case, Eq.~(\ref{EddotnDi})
reads $\ddot{\Delta}_{k}=4\pi G\rho _{0}\Delta _{k}$, implying
exponential growth of $\Delta _{k}$ (the Jeans instability;
see~\cite{har}).

The physical explanation of this rather unexpected stability lies
in the Jeans length associated with the model~\cite{G}. Although
there are always unstable modes (i.e., with wavelength above the
Jeans scale) in a flat space, in a closed universe there is an
upper limit on wavelength. It turns out that, for sufficiently
large speed of sound, all physical wavelengths fall below the
Jeans length. By Eq.~(\ref{es2}), the maximum wavelength $2\pi
a_{0}$ depends on the equation of state and is given by
\begin{equation}
\lambda _{{\rm max}}=\sqrt{\frac{\pi }{G\rho _{0}(1+w)}}\,.
\label{ESlambda}
\end{equation}
The Jeans length is $2\pi a_{0}/n_{{\rm j}}$, where
$\ddot{\Delta}_{k_{{\rm j }}}=0$ and $k_{{\rm j}}^{2}=n_{{\rm
j}}(n_{{\rm j}}+2)$. Hence, by Eq.~(\ref {EddotnDi}), we have
\begin{equation}
\lambda _{{\rm j}}=\left({\frac{c_{{\rm s}}}{\sqrt{4c_{{\rm
s}}^{2}+1}-c_{{\rm s}}} }\right)\lambda _{{\rm max}}\,.
\label{lambdaJ}
\end{equation}
Stability means $\lambda <\lambda _{{\rm j}}$, which leads to
Eq.~(\ref{st}). For dust, $\lambda _{{\rm j}}=0$ and all the modes
are clearly unstable. For radiation, $\lambda _{{\rm
j}}=0.61\lambda _{{\rm max}}$, and for a stiff fluid $\lambda
_{{\rm j}}=0.81\lambda _{{\rm max}}$. In both of these cases the
Jeans length comprises a considerable portion of the size of the
universe, and is greater than all allowed wavelengths. This would
also place important restrictions on the evolution of non-linear
density inhomogeneities by shock damping or black hole formation.

We also note that the stability condition entails neutral
stability; the oscillations in $\Delta$ are not damped by
expansion, since the background is static. In the cosmological
context, where an initial Einstein static state begins to expand
under a homogeneous perturbation, the expansion will damp the
inhomogeneous perturbations. However, the Einstein static will not
be an attractor; instead, the attractor will be de Sitter
spacetime.

Vector perturbations of a fluid are governed by the comoving
dimensionless vorticity $\varpi _{a}=a\omega _{a}$, whose modes
satisfy the propagation equation
\begin{equation}
\dot{\varpi}_{k}=-\left( 1-3c_{{\rm s}}^{2}\right) H\varpi _{k}\,.
\label{dotomi}
\end{equation}
For a fluid Einstein static background, this reduces to
\begin{equation}
\dot{\varpi}_{k}=0\,,
\end{equation}
so that any initial vector perturbations remain frozen. Thus there
is {\em neutral stability against vector perturbations for all
equations of state on all scales.}

Gravitational-wave perturbations of a perfect fluid may be
described in the covariant approach~\cite{c} by the comoving
dimensionless transverse-traceless shear $\Sigma _{ab}=a\sigma
_{ab}$, whose modes satisfy
\begin{eqnarray}
&&\ddot{\Sigma}_{k}+3H\dot{\Sigma}_{k}+\left[
{\frac{k^{2}}{a^{2}}}+2{\frac{K }{a^{2}}}\right.  \nonumber \\
&&~\left. {}-{\frac{8\pi G}{3}}(1+3w)\rho +{\frac{2}{3}}\Lambda
\right] \Sigma _{k}=0\,.
\end{eqnarray}
For the Einstein static background, this becomes
\begin{equation}
\ddot{\Sigma}_{k}+4\pi G\rho _{0}(k^{2}+2)(1+w)\Sigma _{k}=0\,,
\end{equation}
so that there is {\em neutral stability against tensor
perturbations for all equations of state on all scales.} However
this analysis does not cover spatially homogeneous modes. It turns
out that there are various unstable spatially homogeneous
anisotropic modes, for example a Bianchi type~IX mode which we
discuss below. The associated anisotropies can die away in the
case where the instability results in expansion. If this is the
case, despite the extra spatially homogeneous unstable modes, the
expanding inflationary universe will be an attractor.

In summary, for the simplest models (one-component perturbations),
we find neutral stability on all physical inhomogeneous scales
against adiabatic density perturbations if $c_{{\rm
s}}^2>{\frac{1}{5}}$, and against vector and tensor perturbations
for any $c_{{\rm s}}^2$ and $w$, thus generalizing previous
results.

\section{Scalar-field perturbations}

We turn now to the case of a self-interacting scalar field, i.e.,
where the scalar field is governed by a non-flat potential
$V(\phi)$, with initial Einstein static state at $\phi=\phi_0$,
i.e., $V_{0}'=0=\dot{\phi}_{0}$. This is the general dynamical
problem: the stability analysis of the Einstein static universe as
an initial equilibrium position within a physically motivated
(non-flat) potential. (Some realizations of this scenario are
discussed in~\cite{EM}.) Since we are interested only in the
behaviour close to the Einstein static solution, we can treat $H$
and $V'$ as small in the perturbation equations. The lowest-order
solution for density perturbations, $\Delta _{k}^{\z}$, thus
corresponds to $H=0=V'$, and is given by the fluid perturbation
solution, Eq.~(\ref {EddotnDi}),
\begin{eqnarray}
&&\Delta^\z_k(t)=A_k\cos\omega_0 t\,+B_k\sin\omega_0t\,,
\label{lo} \\ &&\omega _{0}^{2}=4\pi G(1+w)\rho _{0}\left[
(k^{2}-3)c_{{\rm s}}^{2}-1 \right] \,,
\end{eqnarray}
with $A_k$ and $B_k$ constants for each mode. The next order
contains contributions from the scalar field perturbations,
\begin{equation}
\Delta _{k}=\Delta _{k}^{\z}+\Delta _{k}^{\f}+\cdots
\,,~~\Delta_{k}^{\f}=\Delta _{\phi \,k}\,,
\end{equation}
during the time when the background is close to the Einstein
static equilibrium position.

The entropy term in Eq.~(\ref{ddotDi}) has no intrinsic fluid
contribution since we assume the fluid is adiabatic. A scalar
field generically has intrinsic entropy perturbations, which
follow from Eq.~(\ref{e}) as~\cite{BED}
\begin{equation}
{\cal E}_{\phi}=
{\frac{1-c_{\phi}^{2}}{w_{\phi}}}\,\Delta_{\phi}\,.
\end{equation}
(See~\cite{btkm} for the corresponding expression in the
metric-based perturbation formalism.) These entropy perturbations
have a stabilizing effect on density perturbations of the scalar
field: the entropy term in Eq.~(\ref{ddotDi}) cancels the
preceding Laplacian term $-c_{\phi }^{2}{\rm D}^{2}\Delta_{\phi}$,
which would produce instability when $c_{\phi }^{2}<0$; what
remains is the term $-{\rm D}^{2}\Delta _{\phi }$, which
contributes to stability.

For an initial Einstein static state, the entropy contribution
from the scalar field to lowest order is,
\begin{equation}
\left[w_\phi{\cal E}_\phi\right]^\f_k=2 \Delta^\f_k\,,
\end{equation}
where we used $[c_\phi^2]^\z=-1$. There is also a relative entropy
contribution, arising from the relative velocity between the field
and the fluid. We expect this contribution to be negligible.

We now expand all the dynamical quantities in Eq.~(\ref{ddotDi})
to incorporate the lowest order effect of the self-interacting
scalar field. In particular, $a = a_0+a_\f$ and
$\rho=\rho_0+\rho_\f$, where $\rho_\f=
{\frac{1}{2}}\dot{\phi}_\f^2+V_\f$, and similarly for the
pressure. Then $w\to w+w_\f $, where
\begin{eqnarray}
w_\f &=&{\frac{p_\f - w\rho_\f }{\rho}}\,,
\end{eqnarray}
and $c_{{\rm s}}^2 \to c_{{\rm s}}^2+ c_\f^2$, where $c_\f^2$ is
determined by the form of the potential $V(\phi)$ near $\phi_0$.
It also follows from the background equations that
\begin{eqnarray}
H_\f^2 &=&{\frac{8\pi G}{3}}\left[{\frac{1}{2}} \dot\phi_\f^2
+V_\f \right]+{ \frac{2 }{a_0^3}}\,a_\f\,, \\ \ddot{\phi}_\f &=&
-V_\f^{\prime}\,.
\end{eqnarray}
Using these results, Eq.~(\ref{ddotDi}) gives the evolution
equation for the contribution from scalar field perturbations:
\begin{equation}  \label{fo}
\ddot{\Delta}^\f_k+\tilde{\omega}_0^2\Delta^\f_k=
F\dot{\Delta}^\z_k+G \Delta^\z_k\,,
\end{equation}
where
\begin{eqnarray}
\tilde{\omega}_0^2 &=& 4\pi G(1+w) \rho_0\left[ (k^{2}-3)(c_{{\rm
s} }^{2}+2)-1\right]\, \\ F(t) &=& -(2-6w+3c_{{\rm s}}^2)H_\f \,,
\\ G(t) &=& {\frac{1}{a_0^2}}\left\{ 12[c_\f^2-w_\f]+24[w-c_{{\rm
s}}^2]{\frac{ a_\f}{a_0}}\right.  \nonumber \\ &&~\left.
+6ww_\f+6c_\f^2-8w_\f \right.  \nonumber \\
&&~\left.{}+[3w^2-8w+6c_{{\rm s}}^2-1]{\frac{\rho_\f }{\rho_0}}
\right. \nonumber \\ &&~\left.{}+k^2\left[c_\f^2-2c_{{\rm
s}}^2{\frac{a_\f }{a_0}}\right]\right\}.
\end{eqnarray}
The change in the frequency, $\omega_0\to \tilde{\omega}_0$, shows
the stabilizing effect of scalar-field entropy perturbations,
which effectively increase the adiabatic sound speed term in
$\omega_0$: $c_{{\rm s}}^2 \to c_{ {\rm s}}^2+2$. The solution of
Eq.~(\ref{fo}) is given by Green's method in the form
\begin{eqnarray}
&&\Delta^\f_k = {\frac{\cos\tilde{\omega}_0t }{\tilde{\omega}_0}}
\,\int\Delta^\z_k\left[
(F\sin\tilde{\omega}_0t)^{\displaystyle{\cdot}}
-G\sin\tilde{\omega}_0t\right]dt  \nonumber \\ &&~~{}-
{\frac{\sin\tilde{\omega}_0t }{\tilde{\omega}_0}}\,\int\Delta^\z_k
\left[ (F\cos\tilde{\omega}_0t)^{\displaystyle{\cdot}}
-G\cos\tilde{\omega} _0t\right]dt\,.
\end{eqnarray}

Since $\tilde{\omega}_{0}^{2}>0,$ for $k^{2}\geq 3$, we see that
stability is not changed by the introduction of scalar-field
perturbations. We note that if general relativity is extended to
include higher-order curvature corrections to the gravitational
Lagrangian, then the existence and stability conditions for the
Einstein static universe are changed in interesting
ways~\cite{BOtt} which are linked to the situation of general
relativity plus a pure scalar field through the conformal
equivalence of the two problems~\cite{BCot}.

\section{Spatially homogeneous tensor perturbations}

We now investigate the stability of the Einstein static universe
to spatially homogeneous gravitational-wave perturbations of the
Bianchi type~IX kind. The anisotropy in these spatially
homogeneous perturbations, which is absent within the Friedmann
family, is what allows for tensor modes.

The Einstein static universe is a particular exact solution of the
Bianchi type~IX, or Mixmaster, universe containing a perfect fluid
and a cosmological constant. The Mixmaster is a spatially
homogeneous closed (compact space sections) universe of the most
general type. It contains the closed isotropic Friedmann universes
as particular subcases when a fluid is present. Physically, the
Mixmaster universe arises from the addition of expansion
anisotropy and 3-curvature anisotropy to the Friedmann universe.
It displays chaotic behaviour on approach to the initial and final
singularities if $w<1$. This is closely linked to the fact that,
despite being a closed universe, its spatial 3-curvature is
negative except when it is close to isotropy.

The diagonal type~IX universe has three expansion scale factors
$a_{i}(t)$, determined by the Einstein equations, which
are~\cite{ll}:
\begin{eqnarray}
\frac{(\dot{a}_{1}a_{2}a_{3})^{\displaystyle{\cdot
}}}{a_{1}a_{2}a_{3}} &=&
\frac{4[(a_{2}^{2}-a_{3}^{2})^{2}-a_{1}^{4}]}{(a_{1}a_{2}a_{3})^{2}}
\nonumber \\ &&~{}+\Lambda +4\pi G(1-w)\rho \,,
\end{eqnarray}
and the two equations obtained by the cyclic interchanges
$a_{1}\rightarrow a_{2}\rightarrow a_{3}\rightarrow a_{1},$
together with the constraint
\begin{equation}
\frac{\ddot{a}_{1}}{a_{1}}+\frac{\ddot{a}_{2}}{a_{2}}+\frac{\ddot{a}_{3}}{
a_{3}}=\Lambda -4\pi G(1+3w)\rho \,,
\end{equation}
and the perfect fluid conservation equation (with $w$ constant)
\begin{equation}
\rho (a_{1}a_{2}a_{3})^{w+1}=\mbox{ const}\,.
\end{equation}
The Einstein static model is the particular solution with all
$a_{i}=a_{0}=$ const. We consider the stability of this static
isotropic solution by linearizing the type~IX equations about it:
$a_{i}(t)=a_{0}+\delta a_{i}(t)$ , $\rho (t)=\rho _{0}+\delta \rho
(t)$, and $\delta p=w\delta \rho $.

The linearized field equations lead to
\begin{eqnarray}
\ddot{\delta a}_{i}+\frac{12}{a_{0}^{2}}\,\delta a_{i}
&=&\frac{8}{a_{0}^{2}}\, \left({\delta a_{1}+\delta a_{2}+\delta
a_{3}}\right) \nonumber \\ &&~{}+4\pi G(1-w)\delta \rho \,.
\end{eqnarray}
The conservation equation to linear order gives
\begin{equation}
\delta \rho =-4\left( {\frac{\Lambda }{8\pi G}}+\rho _{0}\right)
\left( \frac{\delta a_{1}+\delta a_{2}+\delta
a_{3}}{a_{0}}\right).
\end{equation}
If we define an arithmetic-mean perturbed scale factor by
\begin{equation}
\delta A(t)\equiv \frac{\delta a_{1}+\delta a_{2}+\delta
a_{3}}{a_{0}}\,,
\end{equation}
then it obeys
\begin{eqnarray}
\ddot{\delta A} &=&3\delta A\left[
\frac{4}{a_{0}^{2}}-(1-w)(\Lambda +8\pi G\rho _{0}) \right]
\nonumber \\ &=&\frac{3\delta A}{a_{0}^{2}}(1+3w)\,.
\end{eqnarray}
The first term on the right-hand side arises from the pure
spatially homogeneous gravitational wave modes of Bianchi type~IX
(i.e., with $\delta \rho =0=\delta p$). The second term on the
right-hand side arises from the matter perturbations. Thus we see
that the perturbation grows and the Einstein static solution is
unstable to spatially homogeneous pure gravitational-wave
perturbations of Bianchi type~IX,
\begin{equation}
\delta A\propto \exp \left( 2t\sqrt{\frac{3}{a_{0}}}\right).
\end{equation}
When the matter perturbations are included the instability remains
unless $ 1+3w<0$, which we have ruled out for a fluid. In this
case, the perturbations oscillate. This condition corresponds to a
violation of the strong energy condition and this ensures that a
Mixmaster universe containing perfect fluid matter will expand
forever and approach isotropy. Notice that the matter effects
disappear at this order when $w=1$. This is a familiar situation
in anisotropic cosmologies where a $w=1$ fluid behaves on average
like a simple form of anisotropy ``energy".

The instability to spatially homogeneous gravitational wave modes
is not surprising. Mixmaster perturbations allow small distortions
of the Einstein static solution to occur which conserve the volume
but distort the shape. Some directions expand whilst others
contract. Note that these $SO(3)$-invariant homogeneous anisotropy
modes are different to the inhomogeneous modes considered in the
perturbation analysis. Mixmaster oscillatory behaviour is not
picked out by the eigenfunction expansions of perturbations of
Friedmann models.

The relationship between the stability of the Einstein static to
inhomogeneous and spatially homogeneous gravitational-wave
perturbations can be considered in the light of the relationships
between the Bianchi type modes and inhomogeneous modes. In open
universes, Lukash~\cite{luk} has pointed out the correspondence
between Bianchi type~VII anisotropy modes and the inhomogeneous
gravitational wave perturbation spectrum that emerges when
appropriate eignfunctions are chosen for solutions of the Helmholz
equation on negatively curved spaces. These Bianchi type modes
correspond to choosing complex wavenumbers~\cite{l2,b2}. In the
case of a closed universe of Bianchi type~IX a similar
characterisation of the homogeneous tensor mode as arising by
choice of an imaginary wave number would lead to the Mixmaster
instability found above in the case of $k^{2}<-2$. If such modes
are admitted in the spectrum of perturbation modes for the closed
geometry, then they will lead to instability. However, since
superpositions of them lead to non self-adjointness of the
Laplacian~\cite{ms}, there will be problems with quantum analogues
of these modes and it is not clear that they are physically
admissible in the real universe.

\section{Conclusions}

There is considerable interest in the existence of preferred
initial states for the universe and in the existence of stationary
cosmological models. So far this interest has focussed almost
entirely upon the de Sitter universe as a possible initial state,
future attractor, or global stationary state for an eternal
inflationary universe. Of the other two homogeneous spacetimes,
the Einstein static provides an interesting candidate to explore
whether it could play any role in the past evolution of our
Universe. It is important to know whether it can provide a natural
initial state for a past eternal universe, whether it allows the
universe to evolve away from this state, and whether under any
circumstances it can act as an attractor for the very early
evolution of the universe. We might also ask whether it is not
possible for it to provide the globally static background state
for an inhomogeneous eternal universe in which local regions
undergo expansion or contraction, manifesting an instability of
the Einstein static universe. With these questions in mind we have
investigated in detail the situations under which Einstein static
universe is stable and unstable.

We have shown that the Einstein static universe is neutrally
stable against inhomogeneous vector and tensor linear
perturbations, and against scalar density perturbations if $c_{\rm
s}^2>{1\over5}$, extending earlier results of Gibbons for purely
conformal density perturbations. However, we find that spatially
homogeneous gravitational-wave perturbations of the most general
type destabilise a static universe. We pointed out the link that
can be forged between this homogeneous instability and the
behaviour of the inhomogeneous gravitational wave spectrum by
choosing modes with imaginary wave number. Our results show that
if the universe is in a neighbourhood of the Einstein static
solution, it stays in that neighbourhood, but the Einstein static
is not an attractor (because the stability is neutral, with
non-damped oscillations). Expansion away from the static state can
be triggered by a fall in the pressure of the matter. Typically,
expansion away from the static solution will lead to inflation. If
inflation occurs, then perturbations about a Friedmann geometry
will rapidly be driven to zero.  The nonlinear effects (which will
certainly be important in these models because of the initial
infinite time scale envisaged) will be discussed in a further
paper, as will other aspects of the spatially homogeneous
anisotropic modes.

\[ \]
{\bf Acknowledgements}

We thank Marco Bruni, Anthony Challinor and Peter Dunsby for
useful discussions. GE and CT are supported by the NRF. RM is
supported by PPARC, and thanks the Cosmology Group at Cape Town
for hospitality while part of this work was done.

\end{document}